\documentclass[aip,jap,letterpaper,amsmath,amssymb,reprint
]{revtex4-1}
\pdfoutput=1

\usepackage{graphicx}

\begin{document}
\title{Phase separation in Fe$_2$CrSi thin films}
\author{Markus Meinert}
\email{meinert@physik.uni-bielefeld.de}
\author{Torsten H\"ubner}
\author{Jan Schmalhorst}
\author{G\"unter Reiss}
\affiliation{Thin Films and Physics of Nanostructures, Department of Physics, Bielefeld University, D-33501 Bielefeld, Germany}
\author{Elke Arenholz}
\affiliation{Advanced Light Source, Lawrence Berkeley National Laboratory, CA 94720, USA}

\date{\today}

\begin{abstract}
Thin films of a nominal Fe$_2$CrSi alloy have been deposited by magnetron co-sputtering with various heat treatments on MgO and MgAl$_2$O$_4$ substrates. After heat treatment, the films were found to decompose into a nearly epitaxial Fe$_3$Si film with the D0$_3$ structure and Cr$_3$Si precipitates with the A15 structure. We explain the experimental results on the basis of \textit{ab initio} calculations, which reveal that this decomposition is energetically highly favorable.
\end{abstract}

\maketitle

\section{Introduction}
The Heusler compound Fe$_2$CrSi has been predicted to be a half-metallic ferromagnet by numerous theoretical studies based on density functional theory (DFT).\cite{Ishida06, Luo07} However, only few experimental data on this compound are available, such as the lattice constant $a = 5.679$\,\AA{}, the magnetic moment of 2.05\,$\mu_\mathrm{B}$\,/\,f.u., or the Curie temperature of $T_\mathrm{C} = 520$\,K.\cite{Luo07} The compound can eventually crystallize in the A15 phase rather than in the Heusler phase,\cite{Luo07} which was confirmed by recent calculations.\cite{Hamad12} 
While the L2$_1$ structure is an fcc structure (space group Fm$\bar{3}$m) with four atoms per primitive cell, the A15 structure is a simple cubic structure (space group Pm$\bar{3}$n) with eight atoms per primitive cell.

Niculescu \textit{et al.} have studied the solubility of Cr in Fe$_3$Si. They found that Cr is soluble at least up to Fe$_{2.75}$Cr$_{0.25}$Si. At higher Cr content, they observed a decomposition into Cr doped Fe$_3$Si and Cr$_3$Si in the A15 phase.\cite{Niculescu77} 
The binary phase diagram of the Fe-Cr system shows a miscibility gap extending up to at least 950\,K,\cite{Kuwano85} which suggests that also ternary or quaternary phases involving large amounts of Fe and Cr are unlikely to form. The quaternary Co$_2$Cr$_{1-x}$Fe$_x$Al Heusler compounds are a known example of such a system, which shows various spinodal decompositions.\cite{Omar13}
Thus, the Heusler compound Fe$_2$CrSi needs to be synthesized in a non-equilibrium process involving rapid quenching, such as melt-spinning.\cite{Luo07} Thin film growth of an alloy at the nominal Fe$_2$CrSi stoichiometry on a Cr buffer layer has been investigated \cite{Yoshimura08} and attempts to prepare devices like giant magnetoresistance stacks and tunnel magnetoresistance stacks have been made.\cite{Miyawaki11, Wang13}

In this work we investigate magnetron co-sputtered films of a nominal Fe$_2$CrSi alloy. We show that the Heusler phase is unstable and decomposes into Fe$_3$Si and Cr$_3$Si. The experimental results are supported by first principles calculations.

\section{Methods}
The films with a nominal thickness of 20\,nm were deposited by DC and RF magnetron co-sputtering from elemental targets of 99.99\% purity. The stoichiometry was adjusted with the help of x-ray fluorescence measurements to the nominal 2:1:1 composition. MgO and MgAl$_2$O$_4$ (spinel) single crystals with [001] cut were used as substrates. Various heat treatments were applied to the films, including deposition at substrate temperatures of up to 700$^\circ$C and \textit{in situ} post annealing up to 700$^\circ$C after deposition at lower temperature. According to the binary phase diagram of Fe-Cr, higher temperatures are desirable to overcome the miscibility gap, but films deposited at 800$^\circ$C flaked off and could not be investigated. In addition, we have also tested deposition on MgO with a 5\,nm Cr buffer. All films were covered with a 2\,nm MgO film to prevent them from oxidation.

X-ray diffraction (XRD) and reflectivity (XRR) measurements were performed in a Philips X'Pert Pro diffractometer with Cu anode, Bragg-Brentano and point-focus collimator optics and an open Euler cradle.

The films were imaged with atomic force microscopy (AFM) and scanning electron microscopy (SEM). Additionally, energy dispersive x-ray mappings were taken.

Element specific x-ray absorption (XAS) and magnetic circular dichroism (XMCD) measurements were taken at beamline 6.3.1 and 4.0.2 of the Advanced Light Source, Berkeley. The spectra were taken at room temperature. The substrate luminescence was detected with a photodiode to measure the absorption signal of the films.

Formation energies and binary decompositions were investigated with the help of the \textit{Materials Project} database of density functional theory (DFT) calculation results.\cite{materialsproject, materialsprojectonline} High precision calculations of formation energies were done with the full-potential linearized augmented plane-wave (FLAPW) program \textsc{elk}.\cite{elk} All calculations are based on the Perdew-Burke-Ernzerhof (PBE) functional.\cite{pbe}

\section{results}
\begin{figure}[t]
\includegraphics[width=8.6cm]{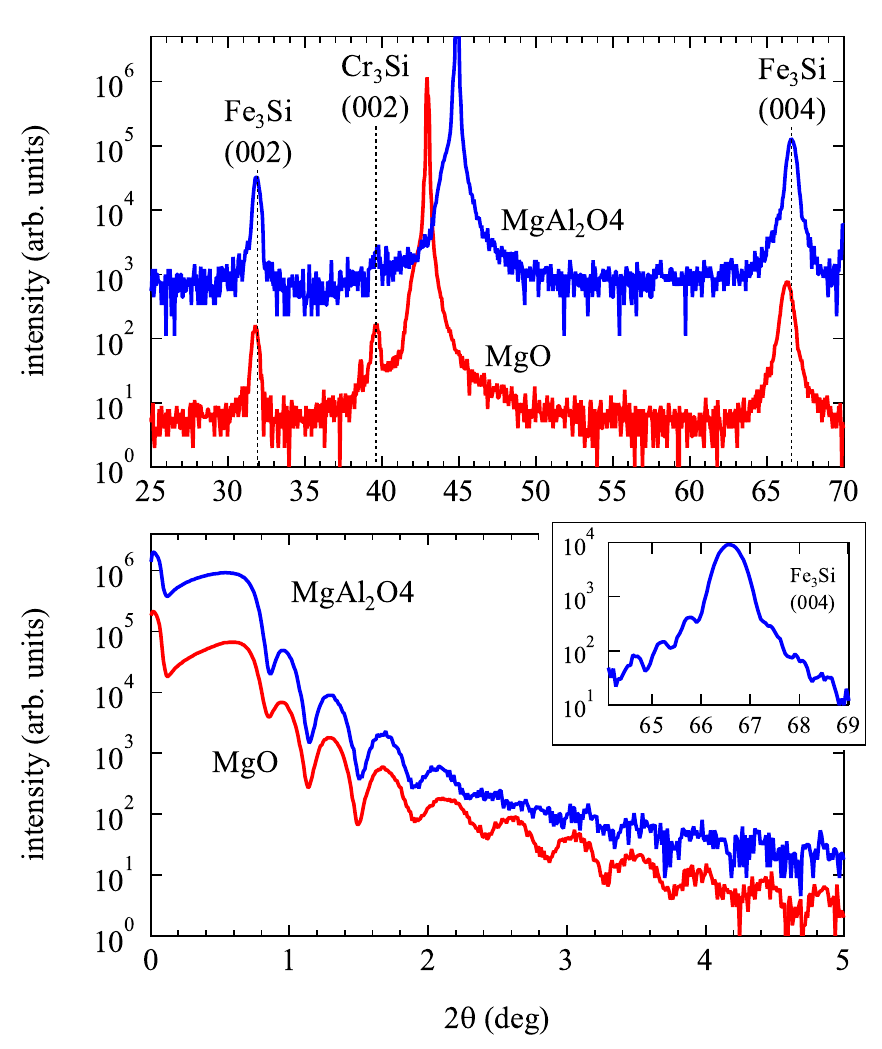}
\caption{\label{XRD} X-ray diffraction (top) and x-ray reflectivity (bottom) spectra of nominal Fe$_2$CrSi films deposited at 200$^\circ$C and \textit{in situ} post-annealed at 700$^\circ$C. The inset shows a high-resolution measurement of the Fe$_3$Si (004) peak of the film deposited on spinel, which exhibits Laue oscillations.}
\end{figure}

The films crystallize after heat treatment on both substrate types. In Fig. \ref{XRD} we exemplarily show two XRD spectra of films deposited on MgO and MgAl$_2$O$_4$ at 200$^\circ$C and \textit{in situ} post-annealed at 700$^\circ$C. The XRD patterns of films with other heat treatments are comparable to those shown. In addition to the expected (002) and (004) peaks of a possible Heusler phase with $a = 5.679$\,\AA{}, we find in most cases a reflection at $2\theta \approx 39.65^\circ$, which can be indexed as a (002) reflection of an A15 phase with $a = 4.545$\,\AA{}. This may be due to a mixture of Fe$_2$CrSi in L2$_1$ and A15 structure or due to precipitate formation. All crystalline films exhibit the (111) reflection which indicates the presence of the L2$_1$ superstructure. The order parameters $S_\mathrm{B2}$ and $S_\mathrm{L2_1}$ are close to unity for the best samples. They were extracted from the measured and calculated intensities of fundamental ((004), (444)) and superstructure ((002) and (222) for $S_\mathrm{B2}$, (111) and (113) for $S_\mathrm{L2_1}$) reflections with Takamura's method.\cite{Takamura09}

Considering the XRR measurements, the films appear to have a smooth surface. Furthermore, the films deposited on MgAl$_2$O$_4$ have extremely narrow rocking curves (measured on the (004) reflection) of 0.03$^\circ$, which is probably still limited by the divergence of the instrument. This points to a very small dislocation density and laterally large grains in the films. The diffraction peaks exhibit Laue oscillations (see inset in Fig. \ref{XRD}), which indicate coherent film growth with smooth interfaces. The crystallite size extracted from the peak width with Scherrer's equation agrees well with the film thickness obtained from the reflectivity curves and supports the picture of coherent film growth.

Surprisingly, AFM topographic images reveal a very rough, grainy surface structure. A typical picture is shown in Fig. 2, which reveals the formation of small (about 100\,nm in diameter) and large (several micrometers) islands. These have been further investigated by EDX analysis in a SEM. The islands are composed of mainly Cr and Si, whereas the flat region in between the small islands is composed of Fe, Si and a small amount of Cr. This finding is illustrated by the EDX mapping in Fig. 3. In conclusion, the Fe$_2$CrSi compound is unstable upon heat treatment and decomposes into other phases.

\begin{figure}[t]
\includegraphics[width=7.6cm]{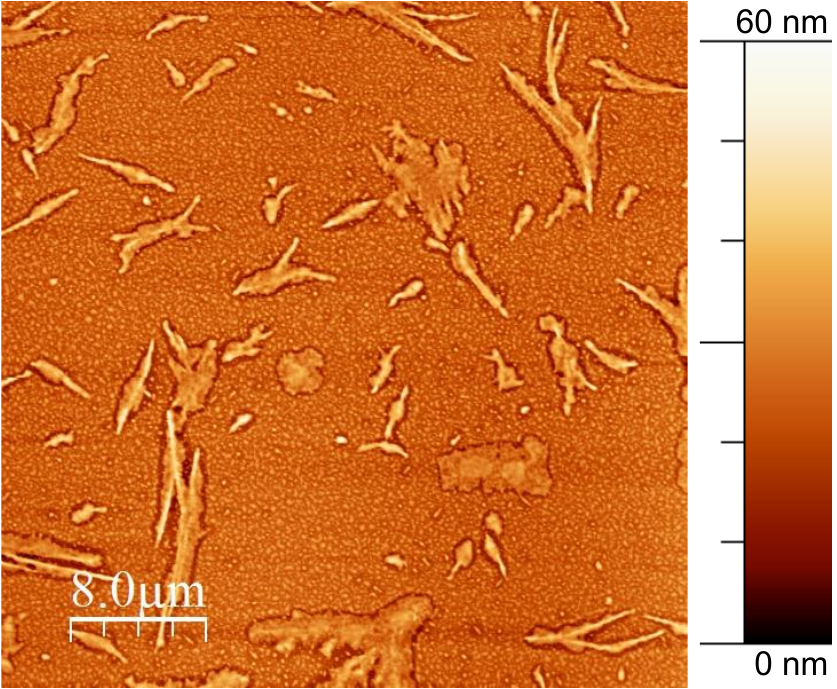}
\caption{\label{AFM} AFM topographic image of an Fe$_2$CrSi film deposited on spinel at 300$^\circ$C and \textit{in situ} post-annealed at 700$^\circ$C.}
\end{figure}

\begin{figure}[b]
\includegraphics[width=8.6cm]{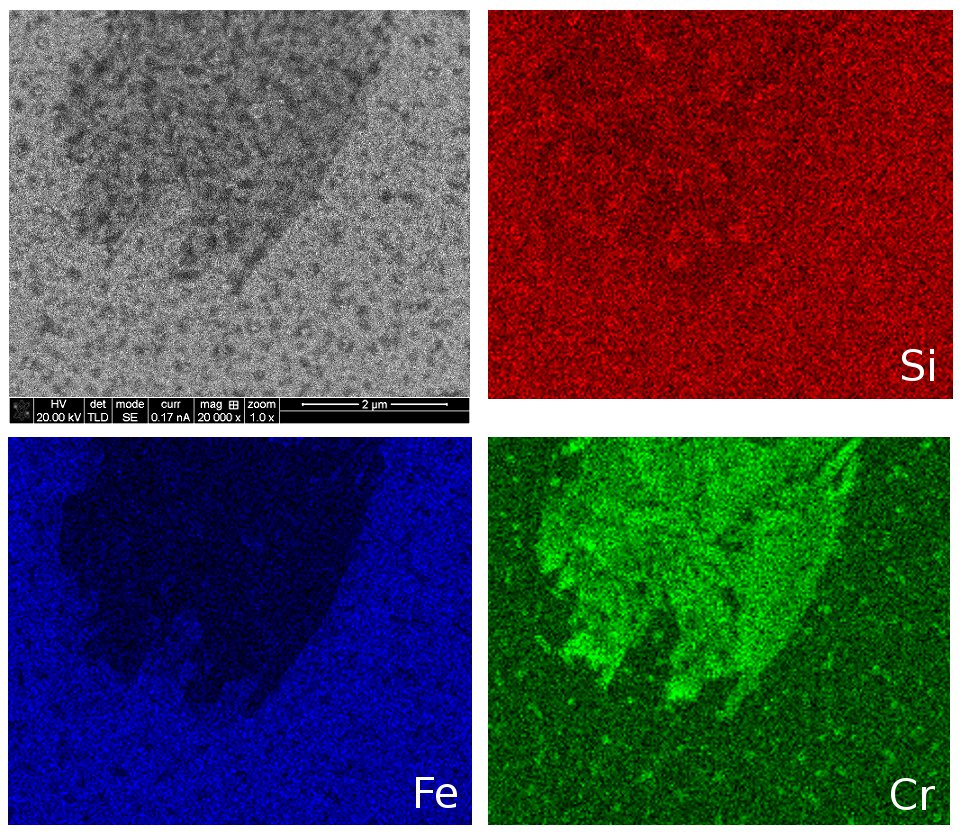}
\caption{\label{EDX} EDX mapping of nominal Fe$_2$CrSi films deposited on spinel at 200$^\circ$C and \textit{in situ} post-annealed at 700$^\circ$C. The field of view is $6.3 \times 5.3$ $\mu$m$^2$ large. The Fe-Cr decomposition is clearly visible.}
\end{figure}

Investigating the \textit{Materials Project} database for possible binary decompositions of the alloy, we find that the reaction 
\begin{equation}
3(2\mathrm{Fe} + \mathrm{Cr} + \mathrm{Si}) \rightarrow 2 \mathrm{Fe}_3\mathrm{Si} + \mathrm{Cr}_3\mathrm{Si}
\end{equation}
is highly favorable. We found the formation energy of the Fe$_2$CrSi compound with L2$_1$ structure to be $-0.93$\,eV/f.u. ($-0.98$\,eV/f.u. for the A15 structure), whereas the formation energy for the above reaction is $-1.31$\,eV/f.u. Thus, the system gains 0.38\,eV/f.u. by decomposing Fe$_2$CrSi into Fe$_3$Si (D0$_3$ structure, $a = 5.654$\,\AA{})\cite{Luo07} and Cr$_3$Si (A15 structure, $a = 4.560$\,\AA{})\cite{Jurisch79}. Indeed, the two lattice constants we have measured are very close to the values of the two binary compounds.

This leads us to the following structural model of the films grown at high temperature: Fe$_3$Si grows epitaxially in the D0$_3$ phase on the two substrate types and forms micrometers large grains. Some Cr is dissolved in the Fe$_3$Si. Cr$_3$Si forms as a precipitate within and on top of the Cr-doped Fe$_3$Si film.

Several samples with different heat treatments have been investigated with XMCD. Most of the samples have very similar properties and we analyze one that has been deposited on MgO at 300$^\circ$C and \textit{in situ} post-annealed at 700$^\circ$C in detail. From the element specific hysteresis loops with the magnetic field perpendicular to the film, we see that Fe and Cr have antiparallel magnetic moments. However, the XMCD signal of the Cr atoms is nearly zero. The saturation field is $(0.7 \pm 0.03)$\,T for both species, which corresponds to a magnetization of $m_\mathrm{tot}^\mathrm{sat} = (2.75 \pm 0.1)\,\mu_\mathrm{B}$ per primitive unit cell, assuming a demagnetization factor of 1. No sign of a second magnetic phase was found (e.g. Fe-doped Cr$_3$Si). The sum rule analysis of the spectra gives an average magnetic spin moment of $m_\mathrm{Fe}^\mathrm{XMCD} = (1.2 \pm 0.1)\,\mu_\mathrm{B}$ for the Fe atoms. This is significantly less than one would expect for pure Fe$_3$Si with 5.5\,$\mu_\mathrm{B}$ per primitive cell and an average magnetic moment of 1.83\,$\mu_\mathrm{B}$ per Fe atom. In any case, this situation is completely different from what is expected for Fe$_2$CrSi in the L2$_1$ structure. Here one expects $0.16\,\mu_\mathrm{B}$ for Fe and $1.65\,\mu_\mathrm{B}$ for Cr.

\begin{figure}[t]
\includegraphics[width=8.6cm]{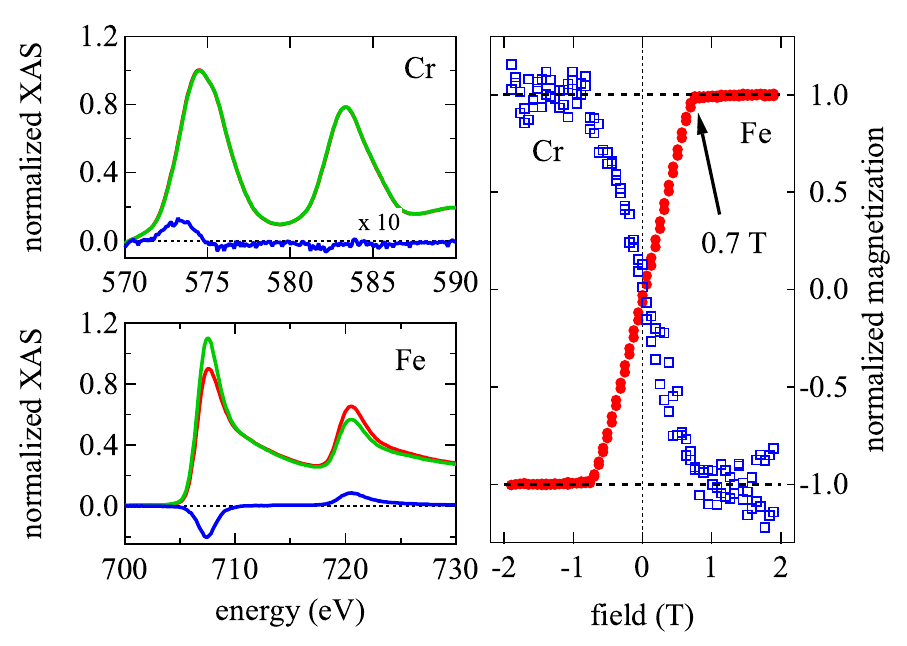}
\caption{\label{XMCD} Left: XAS and XMCD spectra of Cr and Fe of a sample deposited on MgO at 300$^\circ$C and \textit{in situ} post-annealed at 700$^\circ$C. Right: Element specific hysteresis loops of the same sample. All spectra taken at room temperature at BL6.3.1 of the Advanced Light Source. }
\end{figure}

As suggested by Niculescu \textit{et al.},\cite{Niculescu77} we consider a small amount of Cr doped into the Fe$_3$Si and attempt to determine the typical Cr concentration. With 16-atom supercell calculations we find that it is less unfavorable for Cr to enter the Fe 4b sites ($+0.41$\,eV\,/\,atom) rather than the 8c sites ($+0.54$\,eV\,/\,atom). A single Cr atom on an Fe 4b site couples antiparallel to the Fe atoms and reduces the magnetic moments of the Fe 8c atoms. With two Cr atoms on neighboring Fe 4b sites, we find parallel coupling of the Cr atoms with the surrounding Fe atoms. Thus, clustering plays a significant role for the alignment of the magnetic moments. For the Fe 8c moments we find an approximately linear dependence on the Cr concentration $x$, with $x$ defined by Fe$_{3-x}$Cr$_x$Si:
\begin{equation}
m_\mathrm{Fe(8c)} \approx (1-x) \cdot 1.35\,\mu_\mathrm{B}.
\end{equation}

The localized magnetic moment of the Fe 4b atoms is essentially unaffected by the Cr concentration and is about 2.6\,$\mu_\mathrm{B}$. At low Cr concentration, the Cr moment is $-1.6\,\mu_\mathrm{B}$. With these assumptions we have two independent ways to determine $x$:
\begin{align}
2m_\mathrm{Fe(8c)} + [1-x] \cdot m_\mathrm{Fe(4b)} &= [2 + (1-x)] \cdot m_\mathrm{Fe}^\mathrm{XMCD}  \nonumber \\
[2 + (1-x)] \cdot m_\mathrm{Fe}^\mathrm{XMCD} &+ x \cdot m_\mathrm{Cr(4b)} = m_\mathrm{tot}^\mathrm{sat} \nonumber
\end{align}

The first equation is based on the theoretical assumptions on the Fe moments and the measured average Fe moment. The second equation is based on the theoretical assumption on the Cr moment and on the measured saturation magnetization and average Fe moment. We use the low-temperature values for $m_\mathrm{tot}^\mathrm{sat}$ and $m_\mathrm{Fe}^\mathrm{XMCD}$, which are 10\% higher than the room temperature values. This is compatible with a Curie temperature between 500\,K and 600\,K. The equations give $x = 0.33 \pm 0.07$ and $x = 0.32 \pm 0.11$, respectively. These rather large values of $x$ show that the limit of low Cr concentration is not fulfilled. Some Cr atoms are isolated, having negative magnetic moment, while others cluster together, giving a positive contribution to the total signal. On average, a small negative moment remains, in agreement with the small observed negative Cr moment. Thus, the second equation for determining $x$ is less accurate than the first one in addition to the larger experimental uncertainty given above. However, other samples were found to have lower magnetization, thus a higher number of Cr atoms may enter the Fe$_3$Si.

\section{Conclusions}
We have shown that magnetron sputter deposited Fe$_2$CrSi films do not form the desirable L2$_1$ phase. The alloy decomposes into Fe$_3$Si and Cr$_3$Si, which has been shown to be energetically highly favorable. Some Cr may enter the Fe$_3$Si and leads to a significant reduction of its magnetization.

The Fe$_3$Si/Cr$_3$Si mixture may easily be confused with a mixture of Fe$_2$CrSi in L2$_1$ and A15 phases. This can not be distinguished by x-ray diffraction. However, x-ray magnetic circular dichroism offers a simple way to distinguish between the two. In the first case, Fe has a large magnetic moment, while Cr has none. Cr dissolved in the Fe$_3$Si couples antiparallel to the surrounding Fe. In the second case, only the Fe in the A15 phase has a magnetic moment, while Cr has a large moment in the L2$_1$ phase, which should be parallel to the Fe moment of the A15 phase. For all sample types we have investigated, we found large Fe moments and weak antiparallel Cr moments independent of the processing condition, substrates or buffer layers, indicating decomposition.

Due to its instability, the Fe$_2$CrSi compound is probably unsuitable for the proposed spintronic applications.

\acknowledgments
Financial support by the Deutsche Forschungsgemeinschaft (DFG) is gratefully acknowledged. We thank Dr. Karsten Rott for help with EDX mappings, Alexander B\"ohnke for AFM measurements, and Manuel Glas for XAS/XMCD measurements. The Advanced Light Source is supported by the Director, Office of Science, Office of Basic Energy Sciences, of the U.S. Department of Energy under Contract No. DE-AC02-05CH11231.


\begin{thebibliography}{50}
\bibitem{Ishida06} S. Ishida, S. Mizutani, S. Fujii, and S. Asano, Mater. Trans. {\bf 47}, 464 (2006).
\bibitem{Luo07} H. Luo, Z. Zhu, L. Ma, S. Xu, H. Liu, J. Qu, Y. Li, and G. Wu, J. Phys. D: Appl. Phys. {\bf 40}, 7121 (2007).
\bibitem{Hamad12} B. Hamad, J. Khalifeh, Q. M. Hu, and C. Demangeat, J. Mater. Sci. {\bf 47}, 797 (2012).

\bibitem{Niculescu77} V. Niculescu, T. J. Burch, K. Raj, and J. I. Budnick, J. Magn. Magn. Mater. {\bf 5}, 60 (1977).
\bibitem{Kuwano85} H. Kuwano, Trans. Jpn. Inst. Met. {\bf 26}, 473 (1985).
\bibitem{Omar13} A. Omar, M. Dimitrakopoulou, C. G. F. Blum, H. Wendrock, S. Rodan, S. Hampel, W. Löser, B. Büchner, and S. Wurmehl, Cryst. Growth Des., DOI: 10.1021/cg4006136.


\bibitem{Yoshimura08} S. Yoshimura, H. Asano, Y. Nakamura, K. Yamaji, Y. Takeda, M. Matsui, S. Ishida, Y. Nozaki, and K. Matsuyama, J. Appl. Phys. {\bf 103}, 07D716 (2008).

\bibitem{Miyawaki11} T. Miyawaki, K. Takahashi, N. Fukatani, K. Ueda, and H. Asano, IEEE Trans. Magn. {\bf 47}, 2643 (2011).
\bibitem{Wang13} Y. P. Wang, G. C. Han, H. Lu, J. Qiu, Q. J. Yap, R. Ji, and K. L. Teo, J. Appl. Phys. {\bf 114}, 013910 (2013).


\bibitem{materialsproject} A. Jain, G. Hautier, C. Moore, S. P. Ong, C. Fischer, T. Mueller, K. Persson, and G. Ceder, Com. Mat. Sci. {\bf 50}, 2295 (2011).
\bibitem{materialsprojectonline} S. P. Ong, A. Jain, G. Hautier, M. Kocher, S. Cholia, D. Gunter, D. Bailey, D. Skinner, K. Persson, and G. Ceder, http://materialsproject.org/

\bibitem{elk} http://elk.sourceforge.net, Version 1.4.22

\bibitem{pbe} J. P. Perdew, K. Burke, and M. Ernzerhof, Phys. Rev. Lett. {\bf 77}, 3865 (1996).

\bibitem{Takamura09} Y. Takamura, R. Nakane, and S. Sugahara, J. Appl. Phys. {\bf 105}, 07B109 (2009).

\bibitem{Jurisch79} M. Jurisch and G. Behr, Acta Physica Academiae Scientiarum Hungaricae {\bf 47}, 201 (1979).




\end{thebibliography}
\end{document}